\documentclass[preprints,article,accept,moreauthors,pdftex]{mdpi}

\firstpage{1} 
\pubvolume{0}
\issuenum{0}
\articlenumber{0}
\pubyear{2020}
\copyrightyear{2020}
\history{Received: 27 March 2020; Accepted: 20 April 2020}

\usepackage{amssymb}
\usepackage{cleveref}
\usepackage[normalem]{ulem}
\newcommand{\new}[1]{{#1}}
\newcommand{\old}[1]{\sout{}}
\newcommand{\hh}{H$_2$}
\newcommand{\hp}{H$^+$}
\newcommand{\apj}{Astrophys.~J.}
\newcommand{\apjl}{Astrophys. J. Lett.}
\newcommand{\aap}{Astron.~Astrophys.}
\newcommand{\apjs}{Astrophys. J. Suppl.}
\newcommand{\nat}{Nature}
\newcommand{\pasp}{Publ. Astron. Soc. Pac.}

\newcommand{\araa}{Annu. Rev. Astron. Astrophys.}
\newcommand{\aaps}{Astron. Astrophys. Suppl. Ser.}
\newcommand{\jqsrt}{J. Quant. Spectrosc. Radiat. Transf.}
\newcommand{\pra}{Phys.~Rev.~A}

\newcommand{\mnras}{Mon. Not. R. Astron. Soc.}
\newcommand{\jcp}{J. Chem. Phys.}
\newcommand{\physrep}{Phys.Rep.}

\Title{The Leiden Atomic and Molecular Database (LAMDA): Current Status, Recent Updates, and~Future Plans}
\Author{Floris F.S. van der Tak $^{1,2,}$*, Fran\c{c}ois Lique $^3$, Alexandre Faure $^4$, John H. Black $^5$, and Ewine F. van Dishoeck $^{6}$ }
\AuthorNames{Floris van der Tak, Fran\c{c}ois Lique, Alex Faure, John Black, Ewine van Dishoeck}

\address{$^1$ \quad SRON Netherlands Institute for Space Research, Landleven 12, NL-9747AD Groningen, The Netherlands \\
$^2$ \quad Kapteyn Astronomical Institute, University of Groningen, NL-9747AD Groningen, The Netherlands \\
$^3$ \quad Normandie Universit\'e, Universit\'e du Havre \& CNRS, LOMC, F-76063 Le Havre, France; francois.lique@univ-lehavre.fr \\
$^4$ \quad Universit\'e Grenoble-Alpes, CNRS, IPAG, F-38000 Grenoble, France; alexandre.faure@univ-grenoble-alpes.fr \\ 
$^5$ \quad Department of Space, Earth, and Environment, Chalmers University of Technology, Onsala Space Observatory, SE-43992 Onsala, Sweden; john.black@chalmers.se \\
$^6$ \quad Leiden Observatory, Leiden University, 2300 RA Leiden, The Netherlands; ewine@strw.leidenuniv.nl 
}

\corres{Correspondence: f.f.s.van.der.tak@sron.nl; Tel.: +31-50-363-8753 
}

\abstract{The Leiden Atomic and Molecular Database (LAMDA) collects spectroscopic information and collisional rate coefficients for molecules, atoms, and ions of astrophysical and astrochemical interest. 
We describe  the developments of the database since its inception in 2005, and outline our plans for the near future. 
Such a database is constrained both by the nature of its uses and by the availability of accurate data: we suggest ways to improve the synergies among users and suppliers of data. 
We summarize some  recent developments in computation of collisional cross sections and rate coefficients. 
We consider atomic and molecular data that are needed to support astrophysics and astrochemistry with upcoming instruments that operate in the mid- and far-infrared parts of the~spectrum. 
}

\keyword{astronomical data bases; atomic data; molecular data; radiative transfer; ISM: atoms; ISM: molecules; molecular processes; scattering}

\begin{document}

\section{Introduction}

Although baryons constitute only $\approx$5\% of matter in the Universe, the~formation of stars and planets out of baryonic matter is a key astrophysical process.
In the standard model of cosmology, the~Universe expanded, and~matter cooled and recombined to an almost neutral, transparent state without luminosity sources (the 'Dark Ages'), while some initial density fluctuations grew to form structures: clusters, galaxies, and~stars. 
The first generation of stars and the earliest active galactic nuclei re-ionized much of the intergalactic and interstellar gas, into its present ionized transparent state.
Within galaxies, the~cycle of stellar birth, stellar evolution, and~stellar death continues to the present day, \new{with} stellar birth \old{is} accompanied by the formation of planets. 
Nuclear fusion reactions inside stars enrich their composition in heavy elements, which winds and explosions deliver to the interstellar medium. 
The dilute matter in interstellar space plays a crucial role in this cosmic \new{recycling} scheme, both as a source and as a reservoir. 
Astronomers use spectroscopy from radio to X-ray wavelengths to follow the physical and chemical evolution of the interstellar medium (ISM), which is typically in a state far removed from thermodynamic equilibrium. 
Quantitative analysis of atomic and molecular spectra thus requires information about the myriad processes that form and destroy molecules and that redistribute the populations of their internal~states.  

Quantitative analysis of astronomical spectra requires comparison with models. 
In general, a~model of dilute gas consists of (a) a set of coupled differential equations---rate equations---that characterize the populations of the quantum states of the atoms and molecules, and~(b) the equation of radiative transfer that describes the observable intensities of the spectrum in relation to the internal radiative intensities within the source regions. 
These two sets of equations are typically coupled in a nonlinear way. 
In order to perform the analysis (i.e., to solve the equations), it is necessary to assemble large bodies of atomic and molecular data on energies of quantum states, radiative transition probabilities, and~rate coefficients for collision-induced transitions between states. 
These data come from laboratory experiments and quantum mechanical calculations. 
The data are found in publications scattered across the physics, chemistry, and~astronomy literature, and~in various databases---all with a bewildering variety of formats and~units. 

In the limit of very high gas densities, collisional processes predominate and the coupled rate equations can be replaced by the Boltzmann equation, 
\new{since at interstellar temperatures, collisional ionization is negligible. Under~these conditions,} 
the relative populations of all states of an atom or molecule are determined by a single parameter, the~kinetic temperature of the gas $T_{\rm kin}$. 
In the more general case, radiative and collisional processes compete, and~the state populations depend upon temperature, gas density, and~the full spectrum of any internal radiation to which the atom or molecule couples through absorption and stimulated emission. 
Radio astronomers commonly use the term ``Local Thermodynamic Equilibrium (LTE)'' to refer to the high-density limit of molecular excitation. 
This usage ignores the original meaning of LTE in stellar astrophysics, where not only the internal states and molecular motions, but~also the ionization balance, molecular abundances, and~the local source function of radiation at all frequencies are assumed to be in equilibrium at the {\it same} temperature. 
The word ``local'' referred to each interval of depth (or radius) in a stratified atmosphere, where the temperature and total density might vary with depth. 
For clarity, we call the original usage ``strict LTE'', while we use ``quasi-LTE'' for the looser meaning, which is more accurately the assumption of a constant or fixed ``excitation temperature'', $T_{\rm ex}$. 
This excitation temperature is simply the parameter in the Boltzmann \new{distribution}, which generally has a different value for each species in the quasi-LTE case, or~for each transition in the non-LTE case. 
In either case, $T_{\rm ex}$ may deviate substantially from $T_{\rm kin}$. 

The Leiden Atomic and Molecular Database (LAMDA\footnote{\url{https://home.strw.leidenuniv.nl/~moldata/}}$^,$\footnote{Twitter: \url{@lamda\_database}}; \cite{schoeier2005}) aims to provide astronomers with sets of input data for use in spectroscopic models of dilute gas. 
The main application is to the interstellar medium (ISM), notably star- and planet-forming regions, although~applications to comets, (evolved) stellar envelopes, and~dilute (exo)planetary atmospheres are also possible.
The original focus of LAMDA was on rotational transitions of simple molecules, which lie at microwave to far-infrared wavelengths, although~some commonly observed vibrational transitions in the mid-infrared and some atomic fine structure lines were also included.
The database combines two types of data: radiative and collisional (de)excitation rates.
The radiative data alone (term energies, statistical weights, and~spontaneous transition probabilities) are sufficient for the identification of spectral lines, measurements of velocities (via Doppler shifts), and~quasi-LTE estimates of column densities. 
When multiple lines of one species are observed, quasi-LTE models can be used to estimate $T_{\rm ex}$ \cite{goldsmith1999}.
Accurate radiative data are available for most species of astrophysical interest (and many others). 
Data for rotational spectra (microwave, mm-wave,  and~far-infrared) are collected in the
Cologne Database for Molecular Spectroscopy (CDMS;~\cite{mueller2005})\footnote{\url{http://cdms.de}} and Jet Propulsion Laboratory (JPL;~\cite{pickett1998})\footnote{\url{http://spec.jpl.nasa.gov}} molecular databases, 
while the High-resolution Transmission (HITRAN)\footnote{\url{https://www.cfa.harvard.edu/HITRAN/}} molecular absorption database~\cite{rothman2013} extends through near- and mid-infrared wavelengths\old{as well}.
All three of these databases are curated and critically reviewed \new{and the accuracy of radiative data rarely limits the analysis in quantitative astronomical spectroscopy}.  
A useful tool to access the CDMS and JPL databases is the Splatalogue\footnote{\url{http://www.splatalogue.net}}, which allows filtering of data and conversion of units.
Some astrophysical applications at temperatures above $\sim$1000\,K require much more extensive molecular models and line lists.
The ExoMol\footnote{\url{http://www.exomol.com/}} database~\cite{TENNYSON2016} is very useful in this regime, although~caution is needed as the 
data for most molecules are based on theoretical term energies, so~that transition frequencies are generally not of spectroscopic~accuracy.

\begin{table}[H]
\caption{Levels of accuracy in the LAMDA database$^a$. 
}
\label{t:accuracy}
\centering
\begin{tabular}{lll}
\toprule
\textbf{Class}	& \textbf{Accuracy} & \textbf{Type of Calculation} 	\\  
\midrule
A  & $\sim$30\% & Quantum (close coupling or coupled states methods) \\
B & factor of $\sim$2 	& IOS, QCT/statistical calculations; \\
	&				& Born/Coulomb-Born/recoupling approximation \\
C  & factor of $\sim$3--4 	& direct scaling He $\to$ \hh, O $\to$ S \\
D & factor of $\sim$10 	& indirect scalings for similar systems; \\
    &                                	& reduced dimension approaches for reactive systems \\
\bottomrule
\end{tabular}\\
\begin{tabular}{ccc}
\multicolumn{1}{p{\textwidth-.68in}}{\small $^a$: Isotopologues are in the same class as their parents except for H$\to$D substitutions (where the mass correction is $>$30\%), and~symmetry breaking systems (which require their own PES).}
\end{tabular}

\end{table}
The focus of LAMDA is on non-LTE calculations, which provide more accurate estimates of column densities, and~allow determinations of the kinetic temperature and volume density of the gas.
Non-LTE models require cross sections (integrated over velocity into rate coefficients) of the observed species for collisions with the bulk gas component (usually hydrogen in molecular or atomic form). 
For a detailed formulation of the non-LTE problem, see~\cite{vandertak2007}.
For the $\sim$200 molecules which have to date been detected in interstellar space, \new{collisional data} exist for $\sim$50 species (not including atoms), although~for very few molecules have all relevant collision partners been considered.
For inclusion into astrophysical radiative transfer models, these data need to be matched to the available spectroscopic data in their coverage of temperatures and energy levels. 
In addition, existing \new{collisional data} do not always resolve hyperfine (and other) interactions, so that spectroscopy at matching resolution is needed.
Besides this matching of spectroscopy with \new{collisional data}, the~added value of LAMDA lies in its homogeneous data format and its easy access with all species in one~place. 

Users of the LAMDA database are expected to refer to the paper by Sch\"oier~et~al.\ \cite{schoeier2005}.
When individual molecules are considered, references to the original papers providing the spectroscopic and \new{collisional data} must also be made.
The interpretation of astronomical spectra depends on the availability of accurate data; 
conversely, data providers depend on citations and credit for their funding.
References to the database should include the date when LAMDA was consulted, to~make clear which version of the data was used.
In the future, a~more advanced versioning system may be~implemented.

Besides the database, the~LAMDA team also maintains the public version of the widely-used RADEX program for fast non-LTE analysis of interstellar line spectra~\cite{vandertak2007}. 
The program comes in two versions: an online calculator for quick checks, and~a stand-alone version for extensive calculations. 
For publication-quality results, authors should always use the stand-alone version, which gives full control over all input parameters and molecular datafiles.
We encourage users to write a script calling RADEX, to~ensure that their results are~reproducible.

After reviewing current methods to calculate molecular \new{collisional data} (Section~\ref{s:liqueFaure}),
\textls[-15]{this paper describes the actual status of LAMDA (Section~\ref{s:past}) and currently planned updates (Section~\ref{s:present}).
Subsequently, we discuss strategies to deal with missing \new{collisional data} (Section~\ref{s:strategy}) and the molecular data needs for the 2020s (Section~\ref{s:future}).}
The paper focuses on developments since our previous review~\cite{vandertak2011}.
Recent advances in atomic and molecular spectroscopy are discussed by~\cite{schlemmer2019,widicus2019}. 
The theory behind calculations of collision cross-sections is reviewed by~\cite{roueff2013,dagdigian2019}.
For a primer about estimating molecular column densities from observational spectra, see~\cite{shirley2015}. 
Analysis of X-ray spectra is beyond the scope of LAMDA, but~SPEX is a well-known tool for this~\cite{kaastra1996}.
A complementary database for molecular \new{collisional data} is BASECOL~\cite{dubernet2013}\footnote{\url{https://basecol.vamdc.eu/}}.
\begin{table}[H]
\caption{Current entries in the LAMDA database: Atoms and~ions. 
A * symbol in the last column indicates the availability of improved data.
}
\label{t:atoms}
\centering
\scalebox{.9}[.9]{\begin{tabular}{lllll}
\toprule
\textbf{Species} 	& \textbf{Collider} & \textbf{Temperature Range (K)} 	& \textbf{Reference}	& \textbf{Quality Level} \\ 
\midrule
C		& H $^a$	& 10--200		& \cite{launay1977} 		& B* \\ 
C		& e		& 10--20,000	& \cite{johnson1987} 	& B \\
C		& \hp\	& 100--2000	& \cite{roueff1990} 		& B \\
C		& He		& 10--150		& \cite{staemmler1991} 	& B \\
C		& o/p \hh\	& 10--1200	& \cite{schroeder1991} 	& B \\
C$^+$	& H		& 20--2000	& \cite{barinovs2005} 	& A \\ 
C$^+$	& e		& 10--20,000	& \cite{wilson2002} 		& A \\
C$^+$	& \hh\	& 10--500		& \cite{lique2013,wiesenfeld2014} & A \\
N$^+$	& e		& 500--100,000	& \cite{tayal2011} 		& A \\ 
O		& H		& 10--8000	& \cite{lique2018} 		& A \\ 
O		& e		& 50--3000	& \cite{bell1998} 		& A \\
O		& o/p \hh\	& 10--8000	& \cite{lique2018} 		& A \\
O		& He		& 10--8000	& \cite{lique2018} 		& A \\
O		& \hp\	& 10--8000	& \cite{spirko2003} 		& A \\
\bottomrule
\multicolumn{5}{l}{$^a$: More recent calculations exist~\cite{abrahamsson2007}, but~the difference with the posted rates is small.} \\ 
\end{tabular}}
\end{table}
\section{Potential Energy Surface and Scattering~Calculations}
\label{s:liqueFaure}

The theoretical study of collision-induced rovibrational energy
transfer has received much attention over the past 50
years. Astrophysical applications played a significant role in the
development of this research field. Excitation studies of
interstellar molecules began in the early 1970s following the
establishment of \old{the} quantum scattering theory. However, the~computational resources available at that time essentially limited
these studies and most of the \new{collisional data} available were quite
approximative. In~addition, the~helium atom was most of the time used
as a template to mimic H$_2$ collisions. However, over~the last two
decades, the~huge increase of computational resources and the
significant progress in ab initio quantum chemistry have led to
a spectacular improvement of accuracy and to a large variety of
studied systems. From~the direct comparisons between theoretical and
experimental state-to-state collisional cross sections, the~typical
accuracy of modern calculations is 10--20\%
\cite{Yang:11,Chefdeville:13,Chefdeville:15,vogels18}, even for small
polyatomic molecules colliding with H$_2$. Such a good accuracy is
required to guarantee high confidence in radiative transfer
calculations and to \new{allow} a robust determination of molecular column
densities and physical~conditions.

The computation of collisional rate coefficients takes place within
the Born--Oppenheimer approximation for the separation of electronic
and nuclear motions. Scattering cross sections are thus obtained by
solving for the motion of the nuclei on an electronic potential energy
surface. This corresponds to the fixed-nuclei approximation in
electron--molecule collision studies.
The next two subsections describe the recent advances both in terms of computation of
interaction potentials and scattering~calculations.

\subsection{Potential Energy Surface (PES)}

The theoretical study of any scattering process requires the prior
determination of an interaction potential between the particles
involved. This so-called potential energy surface (PES) must be
accurate since dynamical calculations are very sensitive to the PES
quality, and it is impossible to obtain accurate \new{collisional data} if
the PES is not of high quality. In~practice, the~PES accuracy should
be a fraction of the kinetic energy, i.e.,~about 1~Kelvin for
interstellar applications.  The~most accurate approaches are those
based on ab initio quantum chemistry methods. For~non-reactive
systems and because of the low temperatures in the ISM, the~excitation
process of a molecule due to \new{a} collision with a neutral projectile
relates generally to Van der Waals systems in their electronic ground
state. As~these systems often have the particularity of being
adequately represented by a single electronic configuration, the~use
of mono-configurational ab initio methods like the coupled-cluster
methods~\cite{Hampel:92} are preferably used for the determination of
the PESs. The~coupled-cluster method with single, double, and~perturbative triple substitution [CCSD(T)] is considered as the
gold-standard~\cite{vogels18,Kodrycka:19}. The~explicitly-correlated CCSD(T)-F12
approaches~\cite{Adler:07} are nowadays the methods of choice for PES
generation since they allow the use of small atomic basis sets while
maintaining high accuracy. \textls[-15]{They were successfully applied to
many-electron collision systems~\cite{lique2010o2,walker2016}. We note
that in rotational excitation studies most of the PESs are computed by
freezing the molecules \new{in} their vibrationally averaged
geometries. This so-called rigid-rotor approximation was recently
validated against full-dimensional scattering calculations for the
CO--H$_2$ system~\cite{faure16}.} Flexible intermolecular potentials
are thus required \new{only} for vibrational excitation~studies.


\new{
For electron--molecule collisions, multi-configurational ab~initio techniques are preferred because many electron collision processes involve electronic excitation of the target. 
A variety of theoretical approaches are thus available for treating such collisions, such as the Schwinger multichannel, the~complex Kohn variational, and the R-matrix methods (see Tennyson and Faure~\cite{tennyson2019} and references therein). 
The UK Molecular {\bf R}--matrix method is one of the most successful~\cite{tennyson10}. 
Initially developed in the 1940s to treat nuclear reactions, the~{\bf R}--matrix approach was successively adapted to the study of electron--atom and electron--molecule collisions in the early 1970s. 
The {\bf R}--matrix method relies on the division of space into an inner and outer region. 
The inner region is designed to be large enough to contain the entire electron density of the molecular target. 
In the inner region, the scattering electron and target electrons are treated as being indistinguishable and all interactions (polarization, exchange, and correlation) are explicitly considered through multi-reference techniques. 
Conversely, in~the outer region, the scattering electron is only affected by the long-range potential and the physics is much simpler. 
A significant advantage of the {\bf R}--matrix approach is that the inner region problem needs to be solved only once and the energy dependence is entirely obtained from the rapid outer-region~calculations.}

\begin{table}[H]
\caption{Current entries in the LAMDA database: Diatomic~species. 
A * symbol in the last column indicates the availability of improved data.
}
\label{t:diatomic}
\centering
\scalebox{.9}[.9]{\begin{tabular}{lllll}
\toprule
\textbf{Molecule}	& \textbf{Collider}	& \textbf{Temperature Range (K)} & \textbf{Reference} & \textbf{Quality Level} \\  
\midrule
CF$^+$  	 & \hh\      & 10--300 	& \cite{ajili2013} 		& C \\ 
CH            & \hh\     & 10--300 	& \cite{dagdigian2018ch} 	& B \\ 
CH             & H        & 10--300	& \cite{dagdigian2018ch} 	& B \\ 
CN            & \hh\       & 5--300    & \cite{lique2010} 		& C \\ 
CN            & e         & 10--1000  & \cite{allison1971} 		& B* \\ 
CN $^a$   & \hh\      & 5--100   & \cite{kalugina2015} 		& A \\ 
CO         & o/p \hh\  & 2--3000  & \cite{yang2010} 		& A \\ 
CS           & \hh\        & 10--300  & \cite{lique2006} 		& C * \\ 
HCl          & \hh\       & 10--300  & \cite{lanza2014} 		& A \\ 
HF            & He       & 10--300   & \cite{reese2005} 		& A \\ 
HF            & e         & 10--1000 & \cite{vandertak2012} 	& B \\
HF            & \hh\     & 10--150   & \cite{guillon2012} 		& A \\ 
NO           & \hh\       & 10--300    & \cite{lique2009} 		& C \\ 
OH           & \hh\      & 15--300 $^b$ & \cite{offer1994} 	& B \\ 
OH$^+$    & e          & 10--1000 & \cite{faure2012,vandertak2013} & B * \\ 
O$_2$     & \hh\         & 5--350  & \cite{lique2010o2} 		& C * \\ 
SiO         & \hh\        & 5--300 $^c$ & \cite{balanca2018} 	& A \\ 
SiS $^d$ & \hh\         & 10--2000  & \cite{dayou2006} 	& D * \\ 
SO          & \hh\        & 60--300   & \cite{lique2006so} 	& C \\ 
\bottomrule
\end{tabular}}\\
\begin{tabular}{ccc}
\multicolumn{1}{p{\textwidth-.68in}}{\small $^a$: Includes hyperfine structure; $^b$: Includes hyperfine structure for $T$ = 15--200\,K; $^c$: Approximate numbers available for $T$ = 5--1000\,K; $^d$: Scaled from SiO.}
\end{tabular}

\end{table}
\subsection{Scattering~Calculations}

The most accurate approach to treat inelastic scattering remains the
close-coupling method~\cite{Arthurs:60}. When combined with a
high-level PES, it provides highly accurate data. However, for~bimolecular collisions involving `complex' organic molecules ($>$4 atoms) or reactive
species such as ions or radicals, the~increasingly large number of
equations to be solved simultaneously in the close-coupling method
prevents its use, even within the coupled-states
approximation. Several methods have been developed over recent years
to circumvent this issue. The~quasi-classical trajectory (QCT) method,
which has been used with success to generate rate coefficients for
rotationally inelastic transitions~\cite{Faure2007h2o,Faure:16b} 
can be used for de-excitation transitions as long as the molecule is linear or has certain symmetry properties (see the discussion in~\cite{Faure2007h2o}).
Elevated temperature does not seem to be required since for HC$_3$N; good agreement with close-coupling calculations was found down to 10\,K~\cite{wernli2007}.

\textls[-15]{More recently, some progress has been made in the development of
alternative quantum methods to treat inelastic scattering. In~particular, the~wave packet--based Multi Configurational
Time-Dependent Hartree} (MCTDH) method~\cite{ndengue:17} and the mixed
quantum-classical theory (MQCT) \cite{semenov:17} have shown \new{promise}
in the calculation of inelastic cross~sections.

Another quantum alternative is provided by the statistical approach to
bimolecular collisions. The~general idea is to assume the formation of
a long-lived intermediate complex during the collision, resulting in a
statistical redistribution of the population of the internal molecular
levels. A~particular class of statistical approaches consists of
adiabatic capture theories, in~which a series of rotationally or
vibrationally adiabatic potentials is constructed using the long-range
anisotropic potential. One~such method is the statistical adiabatic
channel model (SACM) introduced by~\cite{Quack1975}. This method was
recently combined with global PESs and benchmarked against
close-coupling calculations by~\cite{Loreau2018a}.
For systems where the well depth of the interaction potential is
larger than a few 100 cm$^{-1}$, the~statistical method was
found to reproduce the close-coupling results within a factor of 2 up
to room temperature. A~more elaborate (but more CPU time consuming)
statistical treatment was also employed by~\cite{dagdigian17} in the
  case of CH+H.

\new{
\textls[-15]{For electron--molecule collisions, simple dynamical approaches such as the adiabatic-nuclei-rotational }(ANR) approximation (equivalent to
the infinite-order-sudden approximation; e.g.,~\cite{corey1986}) can be used because the electron collision time is shorter than a typical rotational period
down to the near-threshold regime~\cite{faure06}. The~{\bf R}-matrix method combined with the ANR approximation was checked against experiment for 
the rotational cooling of HD$^+$ and the deexcitation theoretical rate coefficients at 10~K were verified to within 30\% \cite{shafir09}. 
The ANR approximation was also checked against full rovibrational multichannel quantum defect theory (MQDT) calculations. 
It was demonstrated that ANR rotational cross-sections are accurate down to threshold for the molecular ions H$_3^+$ \cite{Faure2006}, HD$^+$ \cite{motapon2014}, and HeH$^+$ \cite{curik2017}. 
We note that recent developments include the extension of the ANR approach to open-shell targets, e.g.,~CN~\cite{harrison2013} and OH$^+$ \cite{hamilton2018}. 
A review of recent electron--molecule calculations of astrophysical interest can be found in~\cite{tennyson2019}.}

\begin{table}[H]
\caption{Current entries in the LAMDA database: Triatomic~species. 
A * symbol in the last column indicates the availability of improved data.
}
\label{t:triatomic}
\centering
\scalebox{.9}[.9]{\begin{tabular}{lllll}
\toprule
\textbf{Molecule}	& \textbf{Collider}	& \textbf{Temperature Range (K)} & \textbf{Reference} & \textbf{Quality Level}\\  
\midrule
C$_2$H        	& \hh\       	& 5--100		& \cite{spielfiedel2012} 	& C \\ 
C$_2$H         	& e          	& 10--1000	& \cite{nagy2015} 		& B \\ 
C$_2$H $^a$	& o/p \hh\	& 10--300 		& \cite{dagdigian2018c2h} & A \\ 
CH$_2$          	&  \hh\      	& 15--300		& \cite{dagdigian2018ch2} & C \\ 
HCN         		& \hh\         & 5--500     	& \cite{dumouchel2010} 	& C \\ 
HCN         		& e            & 5--800		& \cite{faure2007} 		& B \\
HCN  (hfs) 	& \hh\        & 5--30		& \cite{hernandez2017} 	& A \\ 
HCO$^+$ 		& \hh\        & 10--400 $^b$ & \cite{flower1999} 		& B, C* \\ 
HCS$^+$ $^c$ & \hh\       	& 10--2000	&  \cite{flower1999} 		& C \\ 
HDO          	& \hh\        & 5--300		& \cite{faure2012hdo} 	& A \\ 
H$_2$O    	& \hh\        & 5--1500      	& \cite{daniel2011} 		& A \\ 
H$_2$O (rovib) & \hh, e   	& 200--5000	& \cite{faure2008aa} 		& D \\ 
H$_2$S $^d$ 	& \hh\       & 5--1500		& \cite{daniel2011} 		& C \\ 
HNC         		& He        & 5--500       	& \cite{dumouchel2010} 	& C * \\ 
N$_2$H$^+$ 	& \hh\         & 5--50 $^b$     & \cite{daniel2005} 		& C \\ 
OCS        		& \hh\          & 10--100 $^e$ & \cite{green1978} 		& D \\ 
SiC$_2$ 		& \hh\         & 25--125        & \cite{chandra2000} 	& B \\ 
SO$_2$  		& \hh\          & 5--50 $^f$   & \cite{balanca2016} 	& A, B \\ 
\bottomrule
\end{tabular}}\\
\begin{tabular}{ccc}
\multicolumn{1}{p{\textwidth-.68in}}{\small $^a$: Includes hyperfine structure; $^b$: Extrapolation to 2000 K available; $^c$: Scaled from HCO$^+$; $^d$: Scaled from H$_2$O; $^e$: Extrapolation to 500 K available; $^f$: Approximate rates available up to 720 K.}
\end{tabular}

\end{table}
\section{Current Status of~LAMDA} 
\label{s:past}

The LAMDA database combines spectroscopic information with \new{collisional data} for atoms and molecules of astrophysical interest.
To ensure quality and accuracy, the~LAMDA database only contains \new{collisional data} which result from quantum mechanical, quasi-classical, or~statistical calculations, and~which have appeared (or are about to appear) in the refereed literature. 
While published radiative data are usually based on laboratory measurements of line frequencies combined with Hamiltonian models, published \new{collisional data} are mostly the result of (quantum-mechanical) calculations, and~experimental tests are the~exception.

When LAMDA was started in the early 2000s, computational capabilities limited most collision calculations to the use of He as partner, rather than \hh, the~most common species in the cold ISM. 
The~workaround was to scale the He rates to \hh, just correcting for the change in reduced mass between the X-He and X-\hh\ systems in the thermal averaging. 
This scaling is reasonably accurate only for para-\hh, which limits its applicability to temperatures $\ll$78\,K, \new{i.e., the point} where the ortho/para ratio of \hh\ \new{drops below} unity.
While calculations with \hh\ have long become the new standard~\cite{vandertak2011}, the~scaling procedure is still sometimes used, although~the spherical symmetry of He and its lower polarizability compared with \hh\ tends to underestimate the long-range interaction~\cite{roueff2013}, especially with ions.
\new{Collisional data} with He are still useful though, as~ignoring He collisions leads to errors in derived column and volume densities of order 20\%, which is the He/\hh\ abundance~ratio.

In recent years, the~field of collisional calculations has grown \new{more sophisticated}, and~the commonly accepted method to calculate intermolecular interaction potentials is now the coupled-cluster (CC) formalism, usually at the CCSD(T)
level and its explicitly correlated variant or better~\cite{vogels18}; see Section~\ref{s:liqueFaure}.
As a result, most calculations now have \hh\ as a collision partner, and~often resolve its ortho ($J$ = 1) and para ($J$ = 0) moieties. 
Second, hyperfine selective calculations are becoming commonplace, which is important especially for species such as HCN~\cite{vandertak2009} and CN~\cite{faure2012}.
Third, older calculations are often limited to low temperatures ($\lesssim$100\,K) and require extrapolation for applicability in warm star-forming regions. 
Modern calculations cover temperatures up to $\sim$300\,K which largely removes the need for extrapolation and its associated~uncertainty.

\textls[-15]{To guide the users, the~LAMDA database distinguishes four levels of accuracy for its \new{collisional data}, which \new{are summarized in} \Cref{t:accuracy}.}
Most modern PESs are based on CCSD(T) calculations or better, so the accuracy is mostly limited by the scattering treatment.
For most symmetry-conserving isotopologues, we recommend to use the \new{collisional data} for the main isotope; 
only for symmetry-breaking systems and for H$\to$D substitutions does the~isotopolog need its own PES, i.e.,~a separate calculation. 
In particular for hydrides, differences can exceed a factor of 2 due to both kinematics and PES effects (see Scribano~et~al.~\cite{scribano2010}). 
For further discussion on this issue, see Section~\ref{s:present}.

\Cref{t:atoms,t:diatomic,t:triatomic,t:large} list the \new{collisional data} which are included in the LAMDA database as of March 2020.
For atoms and ions, the~database includes the far-infrared fine-structure lines of C, C$^+$, O, and~N$^+$.
Most included molecules are di- and triatomic species, but~some larger molecules with 4--6 atoms are also included.
The collision partner is mostly \hh, although~for some species only scaled He rates are available.
In some cases, \new{collisional data} with H and electrons exist, which are useful to interpret observations of diffuse parts of the ISM, and~regions with high radiation fields.
The last columns of \Cref{t:atoms,t:diatomic,t:triatomic,t:large} give the quality labels as defined in \Cref{t:accuracy}.
We add a * symbol to the label if more accurate data are available; see Section~\ref{s:present} for~details.

The LAMDA data format is used by several well-known radiative transfer programs such as RADEX~\cite{vandertak2007}, RATRAN~\cite{hogerheijde2000}, LIME~\cite{brinch2010}, and~DESPOTIC~\cite{krumholz2014}.
The radiative transfer part of the CLOUDY program~\cite{ferland2017} also uses this format.
The format provides a transparent way for astronomers to use atomic and molecular data without expert knowledge of the physical and chemical literature.
However, proper interpretation of astronomical spectra does require basic knowledge of quantum~physics.

\begin{table}[H]
\caption{Current entries in the LAMDA database: Larger~molecules.
A * symbol in the last column indicates the availability of improved data.
}
\label{t:large}
\centering
\scalebox{.9}[.9]{\begin{tabular}{lllll}
\toprule
\textbf{Molecule}	& \textbf{Collider}	& \textbf{Temperature Range (K)} & \textbf{Reference} & \textbf{Quality Level} \\  
\midrule
CH$_3$OH   	& \hh\       & 10--200     & \cite{rabli2010} 		& B \\ 
CH$_3$CN   	& \hh\    	& 20--140 $^a$ & \cite{green1986} 	& D \\ 
C$_3$\hh\ 	& \hh\        & 30--120    & \cite{chandra2000} 	& C* \\ 
HC$_3$N      	& \hh\        & 10--300 $^b$      & \cite{Faure:16b} & A \\ 
H$_2$CO      	& \hh\        & 10--300      & \cite{wiesenfeld2013} & A \\ 
H$_2$CS $^c$ 	& \hh\      & 10--300     & \cite{wiesenfeld2013} 	& C \\ 
H$_3$O$^+$ 	& \hh\      & 100  & \cite{offer1992}  			& D \\ 
HNCO           	& \hh\       & 7--300   & \cite{sahnoun2018} 	& A \\ 
NH$_3$ $^d$ 	& \hh\       & 15--300  & \cite{danby1988} 		& B* \\ 
NH$_2$D      	& \hh\        & 5--300    & \cite{daniel2014} 		& A \\ 
\bottomrule
\end{tabular}}\\
\begin{tabular}{ccc}
\multicolumn{1}{p{\textwidth-.68in}}{\small $^a$: Extrapolation to 500 K available; $^b$: Hyperfine rates available for 10--100 K; $^c$: Scaled from H$_2$CO; $^d$: More recent calculations by~\cite{bouhafs2017} use a more accurate interaction potential, but cover a smaller temperature range.}
\end{tabular}

\end{table}
\subsection{The LAMDA Data~Format}
\label{ss:format}
\textls[-15]{The format of the LAMDA datafiles is straightforward, flexible, and~easy to use (Table~\ref{t:format}).
Lines starting with a ! sign are to inform human readers, and~are not supposed to be read by computer~programs.}

\textls[-15]{Lines 1--2 name the atomic or molecular species, optionally with a reference for the spectrocopic~data.}

Lines 3--4 give the weight of the species in atomic mass units (amu).

Lines 5--6 give the number of energy levels in the datafile (NLEV).

Lines 7 to 7+NLEV list the energy levels: level number (starting with 1), level energy (cm$^{-1}$), and~statistical weight (to calculate partition functions). In~most datafiles, this information is followed by quantum numbers, which is informative for human readers but not required by computer programs. The~levels must be listed in order of increasing energy.
Spectroscopic accuracy is not required for excitation calculations but~is necessary for spectral synthesis applications (Section~\ref{ss:specratran}).

Lines 8+NLEV and 9+NLEV give the number of radiative transitions (NLIN).

Lines 10+NLEV to 10+NLEV+NLIN list the radiative transitions: transition number, upper level, lower level, and~spontaneous decay rate (s$^{-1}$). 
In many data files, these numbers are followed by line frequencies and upper level energies.
These line frequencies are not read by radiative transfer programs, and~do not have to be of spectroscopic accuracy. 
However, spectroscopic frequencies help human readers to identify the lines and to locate errors in the file.
Many authors also add the upper level energy (in K) after the frequency, which again is informative but not~required.

Lines 11+NLEV+NLIN and 12+NLEV+NLIN give the number of collision partners (NPART).

\medskip

This is followed by NPART blocks of \new{collisional data}:

Lines 13+NLEV+NLIN and 14+NLEV+NLIN give the collision partner ID and reference. Valid identifications are: 1 = \hh, 2 = para-\hh, 3 = ortho-\hh, 4 = electrons, 5 = H, 6 = He, 7 = \hp. 

Lines 15+NLEV+NLIN and 16+NLEV+NLIN give the number of transitions for which \new{collisional data} exist (NCOL).

Lines 17+NLEV+NLIN and 18+NLEV+NLIN give the number of temperatures for which \new{collisional data} are given (NTEMP).

Lines 19+NLEV+NLIN and 20+NLEV+NLIN list the NTEMP values of the temperature for which \new{collisional data} are~given.

Lines 21+NLEV+NLIN to 21+NLEV+NLIN+NCOL list the collisional transitions: transition number, upper level, lower level, and rate coefficients (cm$^3$s$^{-1}$) at each temperature.
The data files only tabulate downward collisional rate coefficients. 
It is presumed that detailed balance (microscopic reversibility) applies to the computation of the corresponding upward rate coefficients from the tabulated values. 
Because of the absence of an energy threshold, downward rate coefficients tend to vary less with temperature, so that interpolation and extrapolation are simpler and more~robust.

\medskip

The treatment of \new{collisional data} between the given temperatures and outside the specified temperature range depends on the program. 
RADEX and RATRAN interpolate between rate coefficients in the specified temperature range. 
Outside this range, they assume that the collisional de-excitation rate coefficients are constant with $T$, i.e.,~they use rate 
coefficients specified at the highest $T$ (400~K in the example of Table~\ref{t:format}) also for higher temperatures, 
and similarly at temperatures below the lowest value (10~K in this case) for which rate coefficients were~specified.

\begin{table}[H]
\caption{Example LAMDA datafile: HCO$^+$.}
\label{t:format}
\centering
\scalebox{.9}[.9]{\begin{tabular}{l}
\toprule
! Molecule \\
HCO+ \\
! Molecular weight \\
29.0 \\
! Number of energy levels \\
21 \\
! Level + Energies (cm-1) + Weight + QuantumNumbers (J) \\
1 0.000000000 1.0 0 \\
2 2.975008479 3.0 1 \\
... \\
21 624.269300464 41.0 20 \\
! Number of radiative transitions \\
20 \\
! Trans + Up + Low + EinsteinA (s-1) + Frequency (GHz) \\
1 2 1 4.251e-05 89.18839570 \\
2 3 2 4.081e-04 178.37481404 \\
... \\
20 21 20 4.955e-01 1781.13802857 \\
! Number of collision partners \\
1 \\
! Collisions between \\
1 H2-HCO+ from Flower (1999) \\
! Number of collisional transitions \\
210 \\
! Number of collisional temperatures \\
12 \\
! Collisional temperatures \\
10.0 20.0 30.0 50.0 70.0 100.0 150.0 200.0 250.0 300.0 350.0 400.0 \\
! Transition + Up + Low + CollisionRates (cm3 s-1) \\
1 2 1 2.6e-10 2.3e-10 2.1e-10 2.0e-10 1.9e-10 1.8e-10 2.0e-10 2.2e-10 2.3e-10 2.5e-10 2.7e-10 2.8e-10 \\
2 3 1 1.4e-10 1.2e-10 1.1e-10 1.0e-10 9.2e-11 8.8e-11 8.4e-11 8.2e-11 8.1e-11 8.3e-11 8.1e-11 8.5e-11 \\
... \\
210 21 20 3.7e-10 3.6e-10 3.6e-10 3.5e-10 3.5e-10 3.5e-10 3.8e-10 4.0e-10 4.4e-10 4.7e-10 5.0e-10 5.2e-10 \\
\bottomrule
\end{tabular}}
\end{table}
\subsection{Common Mistakes and How to Avoid~Them}
\old{Although creating LAMDA-style datafiles is not per se difficult, several issues frequently occur for files contributed by colleagues:}
\new{The format of LAMDA datafiles is designed to be straightforward so that users of modeling codes can create such files on their own. This `FAQ' section offers guidance on some issues that frequently occur in the construction and use of datafiles:}

\begin{itemize}[leftmargin=*,labelsep=5mm]
\item Partner IDs are wrong, so that the program uses \new{collisional data} for the wrong partner. See~Section~\ref{ss:format} for the correct IDs.
\item The actual number of lines does not match NLIN, and it is the same thing for NLEV, NTRANS, NPART, and NTEMP. To~check for such mismatches, RADEX has a `debug' option.
\item \textls[-15]{Transitions occur between levels with the same energy. This happens for example if the spectroscopic data have insufficient frequency resolution. The~obvious solution is to use higher resolution data.}
\item Datafiles contain false metastable states. This occurs especially due to incomplete line lists. In~particular, the~tables of energy levels and transitions must be tested to ensure that there are no levels that completely lack radiative transitions to any lower states, aside from true metastable states. Moreover, true metastable states should always be connected to at least one lower state by a tabulated collisional process. Otherwise, the~solution of rate equations might suffer convergence problems, especially when chemical source and sink terms are ignored.
\item \new{collisional data} are practically always more limited in frequency and energy coverage than spectroscopic data, so to match the two, the~spectroscopy needs to be trimmed. Care must be taken for undesired side effects.
\end{itemize}

\section{Planned Updates of~LAMDA}
\label{s:present}

Recently, a number of collisional calculations have appeared in the literature which need incorporation into LAMDA.
This section orders these updates of the database that are currently foreseen in four types.
The ordering of the types roughly corresponds to the urgency of the updates from high to low.
The actual priority for updating specific \new{collisional data} sets depends also on their added value compared to existing data, and~demand from the community, which in turn depends on how often a species (or transition) is observed, and~in what type of environment (which determines which temperatures and collision partners are appropriate).

The first type of planned updates are species which the database currently does not cover at all.
This category includes 
the fine structure lines of S--He~\cite{lique2018SSi} and Si--He~\cite{lique2018SSi};
the noble gas species ArH$^+$--H~\cite{dagdigian2018}, ArH$^+$--e~\cite{hamilton2016}, and~HeH$^+$--H~\cite{desrousseaux2020};
the diatomic ions SH$^+$--e~\cite{hamilton2018} and CH$^+$--He~\mbox{\cite{hammami2009,turpin2010};}
the neutral diatomics \hh--H~\cite{lique2015h2}, NH--He~\cite{dumouchel2012}, PN--He~\cite{tobola2007}, and~PO--He~\cite{lique2018po};
the triaomics N\hh--\hh\ \cite{bouhafs2019} and C$_3$--He~\cite{benabdallah2008};
and the 'large' species HOCO$^+$--He~\cite{hammami2007}, and~CH$_3$OCHO--He~\cite{faure2011,faure2014}.
We also regard SiS--\hh\ \cite{klos2008} in this category, as~its rates were obtained by scaling SiO~data.

The second type of updates are extensions of existing \new{collisional data} to more transitions or higher temperatures. 
This can be extensions to higher rotational levels, addition of hyperfine structure, and~extension to vibrational or even electronic transitions.
This category includes rovibrational CO--\hh~\cite{castro2017}, rovibrational CO--H~\cite{song2015}, HCO$^+$--\hh\ to 500~K~\cite{yazidi2014}, hyperfine N$_2$H$^+$--\hh\ \cite{lique2015}, and~rovibrational SiO--He~\cite{zhang2018}.

The third type of updates are additional collision partners.
This can be the replacement of He with \hh\ as collision partner, or~adding collisions with H atoms, electrons, CO (e.g., \cite{ndengue2015}) and H$_2$O (the latter two for comets).
This category includes HNC--\hh\ \cite{hernandez2017}, CS--\hh\ \cite{denis2018},
HF--H~\cite{desrousseaux2018},
CN--e~\cite{harrison2013}, OH$^+$--He~\cite{gomez2014}, OH$^+$--H~\cite{lique2016}, OH$^+$--e~\cite{hamilton2018}, \hh O--H~\cite{daniel2015}, CO--H~\cite{song2015}, NH$_3$-H~\cite{bouhafs2017}, O$_2$--\hh\ \cite{lique2014}, HCl--H~\cite{lique2017}, C$_3$H$_2$--\hh\ \cite{benkhalifa2019}, and CH$^+$--e~\cite{hamilton2016}.

The fourth type of updates are data for molecular isotopologues, which come in several varieties.
Some isotopic substitutions fundamentally change the molecular symmetry, such as H$_2$D$^+$--\hh\ \cite{hugo2009} and ND$_2$H--\hh\ \cite{daniel2016ammo}. 
The LAMDA database treats such isotopologues as species on their own.
Other substitutions cause a minor change in the molecular symmetry, such as $^{12}$C--$^{13}$C in C$_3$H$_2$; in such cases, the potential energy surface of the interaction is basically unchanged.
Significant effects on collisional properties occur mostly for H$\to$D substitutions, especially for hydrides: DCO$^+$--\hh\ \cite{pagani2012}, ND--He~\cite{dumouchel2012}, ND$_3$--\hh\ \cite{daniel2016ammo}, D$_2$O--\hh\ \cite{faure2012hdo}, and~N$_2$D$^+$--\hh\ \cite{lin2020}.
If isotopic substitution does not change the symmetry of the molecule at all, the~effect on its collisional properties is small ($<$30\%); this category includes $^{13}$CN--\hh\ and C$^{15}$N--\hh\ \cite{flower2015}, and~N$^{15}$NH$^+$--\hh\ \cite{daniel2016n2hp}
This latter category is the least urgent to update because, for symmetry-conserving isotopic substitutions, simple scaling by (reduced) mass is quite~accurate.

\textls[-15]{Isotope-like scaling is also sometimes used for O$\to$S and certain other substitutions~\cite{vandertak2011}, although~quantitative tests do not exist. 
The expected accuracy is no better than for He$\to$\hh\ substitution, as~the change in rotational constants for O$\to$S is much more pronounced than for e.g.,~ $^{12}$C$\to ^{13}$C.}
Recent results for the H$_2$S--\hh\ system~\cite{dagdigian2020} indicate that the propensity rules are quite different to those of water, which may be expected since H$_2$S is a near-oblate asymmetric top while H$_2$O is closer to~prolate.

For symmetry-conserving isotopic substitutions, using \new{collisional data} for the main isotope is often a reasonable approximation, with~$\approx$20\% accuracy, but~exceptions exist.
One case is ND$_3$ where the substitution does not change the symmetry, but~where more states are allowed than in NH$_3$; see~\cite{vandertak2002}.
Another case is H$_2^{18}$O, where some quantum numbers are inverted w.r.t. H$_2$O, so that upward transitions become downward.
This effect occurs especially for high quantum numbers, \new{when} higher order terms in the Hamiltonian overtake the lower orders.
The best known example is the $3_{13}$--$2_{20}$ transition~\cite{vandertak2006}, but~others exist too; for a recent astrophysical study, see~\cite{putaud2019}.

The H$_2$O molecule is prone to special isotopic effects as it is an asymmetric top with $C_{\rm 2V}$ symmetry, which means that all transitions are $b$-type and change both $K$ quantum numbers ($K_a$ and $K_c$).   
As a result, there is a balance between spectroscopic constants that are powers of $J$, i.e.,~$(B+C)/2$, or~of $K$, i.e.,~$A-((B+C)/2)$.  
Both types of constants will become smaller when substituting a heavier atom, but~not equally.  
The $3_{13}$--$2_{20}$ transition is an rP branch, with~$\Delta${\it J} = $-$1 and $\Delta${\it K}$_a$ = +1.   
The $J$-effect and the $K$-effect will be different (and different at each $J$ and $K$), so that for each transition of this type, it depends on which effect is bigger to decide if the frequency goes up or down.  
The usual notion of all transitions going to lower frequency with heavier isotope is generally true for rR branches and rQ branches, as~long as the molecular symmetry is not~affected.

\subsection{Notes on Individual~Cases}

Hyperfine HCN rates: The integer non-zero spin of the $^{14}$N nucleus leads to hyperfine structure in the rotational lines of HCN, which can be resolved for the lowest-$J$ lines towards clouds which have low intrinsic line broadening due to thermal or turbulent processes.
The intensity ratios of the hyperfine components are useful to estimate the optical depth of the line without assumptions on the spatial extent of the emission.
While LTE is often assumed, the~optical depth estimates are more accurate if collisional deexcitation is taken into account.
The first calculations of hyperfine-resolved \new{collisional data} for the HCN--\hh\ system were made by~\cite{benabdallah2012} with reduced dimensionality for the potential energy surface due to computational limitations.
Upgrading the PES to full dimensionality leads to significant changes in the collisional propensities, as~demonstrated by~\cite{hernandez2017}.
See Braine et al. (submitted to A\&A) for further discussion and~tests.

Rovibrational CO: Thanks to the efforts of Castro~et~al.~\cite{castro2017}, there now exists a set of cross sections and rate data for ortho-H$_2$ and para-H$_2$ collisions with CO that is complete for vibrational quanta $v=0 - 5$ and rotational quanta $J\leq 40$. In~addition, Li~et~al.~\cite{li2015} have published a careful re-evaluation of vibration-rotation line lists for nine isotopologues of CO. We are preparing a LAMDA-format data file that incorporates all of these molecular data. 
Furthermore, Song et al.~\cite{song2015} have calculated rate coefficients for the rovibrational excitation of CO in collisions with H atoms. Woitke et al.~\cite{woitke2018} have shown that such collisions are important in the upper layers of protoplanetary~disks.
 
\subsection{Spectroscopic~Updates}
 
Another area \new{for which} LAMDA could be extended is the role of electronically excited species in the ISM.
One example is O, where Krems~et~al.\ \cite{krems2006} have investigated the H + O($^3$P) $\to$ H + O($^1$D) process. 
Even the downward rate coefficients are relatively slow; however, the~de-excitation process is a means for producing kinetically hot H atoms if O($^1$D) is otherwise excited in warm neutral gas. 
In~a~future update of the data file, the~$^1$D and $^1$S states should be included, together with some of the UV resonance lines, and~extended to higher temperature.
For consistency, the~electron and proton collisions should be included too, since they tend to be even more important in controlling the balance between the $^3$P, $^1$D, and~$^1$S terms of the ground configuration of atomic oxygen. 
In addition,  collisions with protons involve both inelastic and charge-transfer~processes. 

The two most recent quantum mechanical studies of the H$^+$ + O $\rightleftarrows$ H + O$^+$ charge transfer process disagree with each other~\cite{stancil1999,spirko2003}.
This process is extremely important in the interstellar medium and planetary ionospheres. 
It affects both the ion chemistry and the fine-structure excitation of O. 
Its low-temperature behavior needs to be known accurately because the fine-structure splittings in the ground term of atomic O are comparable to interstellar temperatures. 
A definitive investigation is urgently~needed.

\section{What to Do If \new{Collisional~Data} Are Missing}
\label{s:strategy}

For some commonly observed molecules, collision rates with \hh\ are still lacking so that scaled He rates are being used.
Examples include OCS, NO, and~CH$_3$CN (although for the latter, calculations are being performed by M. Ben Khalifa). 
The CF$^+$ ion also lacks \hh\ rates but is less often observed; the PES was recently calculated by Desrousseaux~et~al.~\cite{desrousseaux2020cf+}.
Between He and \hh, the~changes in the collisional rate coefficients are usually a factor of 2--4, or~up to an order of magnitude for hydrides or ions. 
The resulting changes in the derived abundances are usually less, but~having \hh\ \new{collisional data} for these species would ensure that their abundances have the same level of quality as other non-LTE~estimates.

In the astrophysical literature, scalings, estimates, or~educated guesses are sometimes used if actual \new{collisional data} are lacking.
An example are the scaled radiative rates for H$_2$D$^+$ introduced in Ref.~\cite{black1990} and used by Ref.~\cite{stark1999} and Ref.~\cite{vandertak2005}; a similar scaling was adopted in Ref.~\cite{vandertak2016} for H$_2$O$^+$. 
Briefly, the~collision rates of the radiatively allowed transitions are approximated as $Q_0 * S_{ij}$, where $Q_0$ is a typical downward rate coefficient and $S_{ij}$ is the normalized line strength out of initial (upper) level $i$ summed over all final (lower) states.
This factor $S_{ij}$ also enters the calculation of the Einstein A coefficient from an observed microwave intensity; see, e.g.,~the CDMS website\footnote{\url{https://cdms.astro.uni-koeln.de/classic/predictions/description.html}} for details.
The choice of $Q_0$ depends on the molecule and its collision partner. 
Species with high dipole moments ($\gtrsim$1\,D), especially ions, should exhibit strong coupling to \hh, suggesting values for $Q_0$ as high as $10^{-10}$\,cm$^3$s$^{-1}$. 
Apolar species ($\mu \ll$1\,D), and~especially neutrals, are expected to couple more weakly to \hh, leading to $\sim$5--10$\times$ lower $Q_0$ values.  
Atomic H as a collision partner tends to show higher rates than \hh, at~least for light (reactive) hydrides, where we recommend high $Q_0$ values like those for \hh\ on a polar neutral.
For heavier neutral species such as NH$_3$ and H$_2$O colliding with H, lower $Q_0$ values are in order, like for \hh\ collisions.

The validity of scaling with $S_{ij}$ critically depends on the dipole selection rule that only $\Delta${\it J}~=~1 transitions are allowed. 
Non-radiative transitions are assumed to have negligible collisional rate coefficients, which is of course a simplification.
For certain cases, this scaling does not work at all: NH$_2$ is an example~\cite{bouhafs2019}.

In the case of inelastic collisions between electrons and molecules, the~Born approximation works well when quantum mechanical calculations are not available (see review by~\cite{itikawa2005}). 
The Born approximation cross sections for rotational transitions can be written in terms of radiative transition probabilities~\cite{Itikawa1969}, as~there is a strong propensity for dipole selection rules. 
For electron--ion collisions, the~Coulomb--Born approximation is used.
Both approximations work best for large dipoles ($\mu \gtrsim 2$\,D); otherwise, the dipole-forbidden transitions have non-negligible rates.
See~\cite{tennyson2019} for further discussion.

To interpret spectroscopic observations of comets and estimate the production rates of various species in the coma, most existing \new{collisional data} are not suitable since the main collision partner in this case is H$_2$O.
For this situation, the `Boltzmann' recipe is often used, where collisions are supposed to redistribute the molecule according to the Boltzmann distribution~\cite{crovisier1983}.
 
Although calculated rates are always preferred over scalings and estimates, astronomers are regularly observing molecules for which no \new{collisional data} exist.
This situation is unlikely to change anytime soon; however, where 10 years ago the gap between theory and observation was widening rather than closing~\cite{vandertak2011}, it seems that today the~rates of new molecular detections in space and of newly computed \new{collisional data} have converged at a few species per year.  
When observing species without \new{collisional data}, guessed collision rates are clearly preferred over guessed excitation temperatures, which are assumed to hold for all transitions of a species.
In turn, guessed excitation temperatures are clearly preferred over quasi-LTE ($T_{\rm ex} = T_{\rm kin}$) which tends to hold only for low-$J$ transitions in regions with high gas densities.
Therefore, we consider scaled \new{collisional data} useful in cases where detailed calculations are infeasible, such as for rovibrational and/or vibronic transitions, especially for large~molecules.

\section{Molecular Data Needs for the~Future\label{s:future}}
\unskip

\subsection{\new{Collisional~Data}}

As discussed in Section~\ref{s:liqueFaure}, it seems now possible to compute highly accurate
collisional rate coefficients for many simple non-reactive species in
collision with electrons, He, H, and H$_2$. Despite two decades of
theoretical efforts, however, there are still some \new{collisional data} to
update. In~particular, molecules for which only the He collisional
partner was considered (e.g., NO, CH$_2$, HCS$^+$, ...) have to be the
object of new scattering studies considering H$_2$ as a~projectile.

Then, the~new methods that have been developed should allow for providing
\new{collisional data} for reactive species such as H$_2$O$^+$. Collisional
excitation studies implying complex organic molecules can also be
studied using approximate scattering approaches such as MQCT. This
method was compared to close-coupling results for the rotational
excitation of methyl formate (HCOOCH$_3$) by He
\citep{faure2014,semenov15}.

Finally, extension of the available data to higher temperatures and to
a larger number of rotationally excited levels is desirable and
sometimes even critical. With~increasing temperature, vibrational
excitation thresholds open and become a new challenge to overcome. In~the last decade, pioneering studies have considered the excitation of
the torsional motion of CH$_3$OH \citep{rabli11} and the bending
motion of triatomics such as HCN \citep{stoecklin13} by He. Recent
\new{studies} have also treated the rovibrational excitation of diatomics by
H$_2$ (see e.g.,~\cite{yang18}). Future work will need to address the
coupling of the vibration(s) of nonlinear polyatomic targets with the
H$_2$ rotation.

\subsubsection{Radio and Far-Infrared~Data}

From an astrophysical point of view, several types of \new{collisional data} are needed to interpret observations from upcoming telescopes.
The first are \new{collisional data} for large organic species. 
Currently, the~largest species in LAMDA are CH$_3$OH and CH$_3$CN, but~molecules with up to 13~atoms have been detected in the ISM\footnote{For an up-to-date overview, see {\url{https://cdms.astro.uni-koeln.de/classic/molecules}}}, with~c-C$_6$H$_5$CN being the current record~\cite{McGuire2018}.
In addition, the~`buckyball' molecules C$_{60}$, C$_{60}^+$ and C$_{70}$ have been detected in circumstellar shells and nebulae~\cite{cami2010}.
Spectral surveys of star-forming regions with ALMA now reveal myriads of lines from organic molecules with $\approx$10 atoms, which \new{are} complex by astrophysical (if not by chemical) standards~\cite{allen2017,bogelund2019}. 
\new{collisional data} are needed to carry out non-LTE analysis for these species, but~which data are needed depends on the situation.
For studies of dark clouds, \new{collisional data} for low-$J$ transitions at low temperatures would be sufficient\old{,} for the species detected in such clouds (e.g., \cite{bacmann2012,agundez2019}). 
For warm and dense star-forming regions, quasi-LTE models give good fits to the data, and~molecular column densities are accurate to $\approx$30\% \cite{jorgensen2018} within reasonable ranges of $T_{\rm ex}$, although~the assumption of a single $T_{\rm ex}$ for all lines may give some additional uncertainty.
Furthermore, to~interpret the inferred temperatures (kinetic or radiative) and to determine the importance of infrared pumping, \new{collisional data} are essential.
Detailed quantum mechanical calculations exist for CH$_3$CHO~\cite{faure2014} and are being calculated for NH$_2$CHO, CH$_3$OCH$_3$, and~CH$_3$CHCH$_2$O~\cite{faure2019}.
For larger molecules, such calculations are currently \new{un}feasible, but~at the temperatures of $\sim$100--300\,K of star-forming regions, approximate recipes and extrapolations for their collision rates would be adequate.
First, steps toward this goal were taken by~\cite{faure2008iau}, and~further worked out by~\cite{faure2008aa} for the rovibrational excitation of H$_2$O.

Second, \new{collisional data} are needed for several small species:

\begin{itemize}[leftmargin=*,labelsep=5mm]
\item HeH$^+$ which is a key species in the chemistry of the early Universe~\cite{galli2013} and recently detected in a planetary nebula~\cite{guesten2019,neufeld2020}. 
\new{Collisional data} with electrons and H atoms exist~\cite{hamilton2016,desrousseaux2020}, but~data with \hh\ are still missing, although~the HeH$^+$ zone contains little \hh.
\item PH$_3$ as a key species in interstellar phosphorus chemistry~\cite{agundez2014,rivilla2020} which also has application to Jupiter and exoplanet atmospheres~\cite{sousa2020}. 
Its structure resembles that of NH$_3$, but~the barrier to inversion (umbrella-mode vibration) is so much higher in PH$_3$ that the inversion splitting is unmeasurably small in the ground state and thus does not lead to any inversion transitions~\cite{sousa2016}. 
Hence, inversion-resolved collision rates for NH$_3$ cannot be applied as such to PH$_3$, although~inversion-resolved rates could be averaged and summed to provide rotational~rates. 
\item H$_2$O$^+$ as a key species in the chemistry of H$_2$O-like ions which is a useful probe of the cosmic-ray ionization rate~\cite{indriolo2015} and for which \new{collisional data} are underway as part of the ERC Consolidator Grant COLLEXISM project (PI F. Lique).
\item HC$_5$N, which is important to study the formation of interstellar carbon chains. Work is underway by Lique and Dawes.
Ideally, a recipe should be developed to extend the calculations to HC$_n$N with $n$~=~7, 9, 11, ...
\end{itemize}

\new{Third,} a recipe for scalable \new{collisional data} for isotopic species would be useful to improve abundance estimates for commonly observed isotopologues such as DCO$^+$ and DCN.
For isotopes of C, N, O, and~heavier atoms, the~relative change in mass is too small to affect the collisional properties of the molecules significantly.
However, simple recipes are only likely to apply to substitutions which conserve molecular symmetry and which do not involve spectral complications (see Section~\ref{s:present}).

Fourth, \new{collisional data} with water as partner are needed to interpret cometary observations. 
Full quantum data for HCN and CO have just appeared~\cite{dubernet2019,faure2020} and more work is underway. 
The difficulty with molecule--H$_2$O collisions is twofold: first, the~potential well is much deeper than in molecule--H$_2$ systems and, second, H$_2$O has a denser rotational spectrum than H$_2$. 
This means that full close-coupling calculations are prohibitively expensive due to the excessively large number of angular couplings. 
\mbox{As a result}, the~data of~\cite{dubernet2019} were obtained using partially converged coupled-states calculations while those of~\cite{faure2020} were computed within the SACM approximation. 
These two pioneering studies have shown, however, that propensity rules are less pronounced than in molecule-H$_2$ collisions and that pair correlations and resonant energy transfer between the target and H$_2$O play a critical role.
We note that, for comets at large heliocentric distances (typically $R_h>3$~au), CO becomes another important or even dominant collider because of its low sublimation~temperature.

\subsubsection{Near- and Mid-Infrared~Data}

The fifth and final type \new{of needed data} are \new{collisional data} for atoms and small molecules to interpret spectra from upcoming mid-infrared observatories, especially JWST/MIRI~\cite{wright2015}, ELT/METIS~\cite{brandl2018}, and~SPICA/SMI~\cite{roelfsema2018}. 
These facilities will probe the physical conditions and the chemical composition of the inner parts of protoplanetary disks.
Calculated rate coefficients already exist for CO rovibrational and \hh\ and HD rotational lines (\Cref{t:atoms,t:diatomic,t:triatomic,t:large}).
In Nijmegen, measurements are ongoing for the $\nu_2$ ('umbrella') mode of NH$_3$ \cite{asselin2018}, and~calculations for the rovibrational transitions of CO$_2$ \cite{bosman2017} and C$_2$H$_2$ (Selim et al., in~prep).
Recently, rovibrational calculations have also been made for H$_2$O--\hh\ \cite{stoecklin2019} and HCN--He~\cite{stoecklin13},
while for species like SO$_2$ and OH that are commonly observed at mid-IR wavelengths, no data for their mid-IR transitions exist yet. 
Particularly important are symmetric molecules without a dipole moment such as CH$_4$ \cite{sahnoun2020} that cannot be observed at (sub)mm wavelengths, \new{but} are key players in the organic chemistry of inner disks~\cite{pontoppidan2010}.
In addition, atomic line data are needed, in~particular for the mid-IR fine structure lines of Ne$^+$, Fe, and~Fe$^+$.
For the ionic species, data exist for collisions with electrons, which tend to dominate the excitation. 
The corresponding data files are in preparation~\cite{johnson1987Ne,ramsbottom2007}.
\new{Likewise, Pelan and Berrington~\cite{pelan1997} have calculated e-impact rates included in the upcoming LAMDA file for Fe, although~their work applies only to the two lowest terms of even parity, viz. a$^5$D and a$^5$F.}
For~collisions \new{of atoms and ions} with H and especially \hh, \new{the calculations are complicated as} reactions compete with inelastic (de)excitation.
 

\subsection{Spectroscopic Data and Radiative Transfer~Tools}
\label{ss:specratran}

Finally, a~number of developments in spectroscopy and radiative transfer are needed for proper interpretation of astronomical spectra from upcoming facilities:

\begin{itemize}[leftmargin=*,labelsep=5mm]

\item The interpretation of laboratory spectra through Hamiltonian models and the creation of synthetic line lists for comparison with observed spectra is commonly done with the programs SPFIT and SPCAT, developed by Herb Pickett. Currently, the~use of these programs requires significant specialization and training. User-friendly versions of these programs are needed to ensure proper interpretation of laboratory spectra in the future. The~PGOPHER\footnote{\url{http://pgopher.chm.bris.ac.uk/}} package~\cite{western2017} is a major step in this~direction.

\item (Sub-)mm laboratory spectra of large/complex organic molecules are needed, including vibrational modes and isotopic species, to~beat line confusion in ALMA spectral surveys such as PILS~\cite{jorgensen2018} and ReMoCa~\cite{belloche2019}.

\item An update to RADEX is in preparation, which is able to include formation/destruction processes in the radiative transfer problem.
Such calculations require state-to-state reactive \new{collisional data}, which exist only for some cases such as CH$^+$ \cite{faure2017} and OH$^+$ \cite{lique2016}.
This new version of RADEX is also capable of multi-molecular spectral synthesis, similar to the programs XCLASS~\cite{moeller2017} and CASSIS\footnote{\url{http://cassis.irap.omp.eu/}}.

\end{itemize}

\section{Conclusions}

In conclusion, it is encouraging to see that, for many species of astrophysical interest, accurate \new{collisional data} are now either existing or underway.
Thanks to 1--2 decades of dedicated work, the~calculation of \new{collisional data} is now keeping pace with the rate of new molecular detections.
With~many updates planned or ongoing, we consider the LAMDA database ready for the 2020~decade.

\vspace{6pt} 

\authorcontributions{
Since 2005, many researchers have contributed to the development of the LAMDA database by providing or testing datafiles. 
In alphabetical order: Arthur Bosman, Christian Brinch, Simon Bruderer, Emmanuel Caux, Paul Dagdigian, Fabien Daniel, David Flower, Asunci\'on Fuente, Adam Ginsburg, Javier Goicoechea, Paul Goldsmith, Vedad Hodzic, Jens Kauffmann, Mher Kazandajin, Maja Kazmierczak, Lars Kristensen, Boy Lankhaar, Zs\'ofia Nagy, David Neufeld, Marco Padovani, Laurent Pagani, Dieter Poelman, Ryan Porter, Megan Reiter, Markus R\"ollig, Evelyne Roueff, Ian Smith, Richard Teague, Wing-Fai Thi, Kalle Torstensson, Charlotte Vastel, Ruud Visser, Kuo-Song Wang, and Matthijs van der~Wiel.
}

\funding{The LAMDA database is supported by the Netherlands Organization for Scientific Research (NWO), the~Netherlands Research School for Astronomy (NOVA), and~the Swedish Research Council.
In addition, F. L. and A.F.  acknowledge the European Research Council
(Consolidator Grant COLLEXISM, grant agreement 811363) and the
Programme National ``Physique et Chimie du Milieu Interstellaire''
(PCMI) of Centre National de la Recherche Scientifique (CNRS)/Institut
National des Sciences de l'Univers (INSU) with Institut de Chimie
(INC)/Institut de Physique (INP) co-funded by Commissariat \`a l'Energie
Atomique (CEA) and Centre National d'Etudes Spatiales
(CNES). F.L. also acknowledges financial support from the
Institut Universitaire de~France.}


\acknowledgments{The authors thank Gerrit Groenenboom, Ad van der Avoird, and Bas van de Meerakker (Nijmegen) for input, 
Evelyne Roueff (Paris) and Laurent Wiesenfeld (Saclay) for comments on the manuscript, and John Pearson (JPL) for useful~discussions.
We also acknowledge our 3 referees for their careful reading of the manuscript.
}

\conflictsofinterest{The authors declare no conflict of~interest.} 

\reftitle{References}

\end{document}